\input{psfig.sty}

  \def\selectedoptions{final}

\documentclass[
   \selectedoptions
  ]
  {aipproc}

\def\selectedlayoutstyle{6x9}
\layoutstyle\selectedlayoutstyle 

\SetInternalRegister\hbadness{8000} 

\newcommand\doingARLO[2][]{%
  \ifx\mmref\undefined #1\else #2\fi
} 
  
\newcommand{\etal}{{\it et al. }}
\begin{document}

\title 
[] {Studying the Accretion Disks in Black Hole X-ray Binaries 
with Monte-Carlo Simulations} 

\author{Yangsen Yao}{
  address={University of Alabama in Huntsville and National Space Science and Technology Center, \\Physics Department, Huntsville, AL 35899, USA},
  email={yaoys@jet.uah.edu},
  homepage={http://jet.uah.edu/~yaoys}
} 

\iftrue 
\author{Shuang Nan Zhang}{
  address={University of Alabama in Huntsville and National Space Science and Technology Center, \\Physics Department, Huntsville, AL 35899, USA},
  email={zhangsn@jet.uah.edu},
  homepage={http://jet.uah.edu/~zhangsn}
} 
\author{Xiaoling Zhang}{
  address={University of Alabama in Huntsville and National Space Science and Technology Center, \\Physics Department, Huntsville, AL 35899, USA},
  email={yaoys@jet.uah.edu},
  homepage={http://jet.uah.edu/~xizhang}
} 

\fi 

\copyrightyear  {2001} 

\begin{abstract}
Understanding the properties of the hot corona is very important for 
studying the accretion disks in black hole X-ray binary systems. Using the 
Monte-Carlo technique to simulate the inverse Compton 
 scattering process between disk photons and electrons in the hot corona, 
we have produced two table models in the XSPEC package. 
Applying the models to the broad-band BeppoSAX observations of the black hole candidate XTE J2012+381, 
we demonstrate the power of this table model. Our results indicate that the electron distribution 
in the corona has a powerlaw shape with a spectral index around 4 and the 
size of the corona is just several tens of gravitational radius and the inclination 
angle of the disk is around 60 degrees.
\end{abstract}

\date{\today}

\maketitle 

\section{Introduction}
\vspace{-3mm} The spectrum of a black hole X-ray binary system usually 
consists of two components (see Zhang, et al. 1997 for a review and references therein): 
blackbody-like component and powerlaw-like component. The blackbody-like 
component turns off above 20 keV and is believed to be emitted from the 
accretion disk. The powerlaw-like component can extend up to 200 keV and is 
believed to be produced in a hot corona around the black hole through the 
inverse Compton scattering process between the disk photons and electrons 
in the corona. This process can be described by the Kompaneets's equation, 
which is a non-linear diffusion equation and very difficult for 
analytical solutions. 

Analytic solutions have been worked out by 
Sunyaev and Titarchuk (1980) and Titarchuck (1994) based on these assumptions: 1) uniform corona with either 
spherical or disk geometry; 2) seed photons are emitted in the center of 
the corona; 3) the seed photon distribution follows the Wien form; 4) the 
thermal temperature of electrons in the corona is much higher than the Wien 
temperature of the seed photons. However, some or all of these assumptions are 
not appropriate for black hole binary systems. 

In the X-ray astronomy community, the multi-color blackbody (Mitsuda et al., 1984; 
Makishima et al., 1986; {\em diskbb} in {\em XSPEC}) 
plus a powerlaw model has been employed traditionally to fit the energy 
spectra of the black hole X-ray binary systems; this model is non-physical 
and unreasonable as we pointed out in our previous paper (Zhang et al., 
2001). 

In this paper, we use the Monte-Carlo technique to simulate the inverse 
Compton process in the corona; two powerful table models (as the standard 
model in $XSPEC$ package) have been built up based on our simulation 
results. Using these table models to fit the data on XTE J2012+381 with 
BeppoSAX, we estimated several physical parameters of the accretion disk 
and corona. 
\section{Monte-Carlo simulations}
\vspace{-3mm} In our simulations, we assume that the accretion disk is the 
Keplerian disk, and during the accretion process, the potential energy loss 
of the accreted material is radiated away in blackbody radiation. 
Therefore, the temperature distribution along the accretion disk is $T 
\propto r^{-3/4}$. The corona may take either spherical or disk 
geometry; the electron density distribution in the corona may be either 
uniform or non-uniform (take the powerlaw for non-uniform distribution); 
and the electron energy distribution in the corona may take either thermal form 
or non-thermal form (powerlaw). 

Our simulation results can be summarized as following: 1) for a given 
electron temperature, the scattered photon spectrum in the high energy band is 
determined by the optical depth of the corona (FIGURE 1.a); 2) the shape of 
scattered photon spectrum in the low energy band is related to the 
geometrical size of the corona (FIGURE 1.b); 3) for the same properties of 
disk and corona, the ratio between the flux of soft component and that 
of the hard component is related to the inclination angle of the accretion disk 
(FIGURE 1.c); 4) the high energy portions of the spectra are very different 
for different electron energy distributions in the corona, even though the 
low energy portions are quite similar (FIGURE 1.d).

Based on our simulation results, we have built up two 
different table models for thermal and powerlaw electron energy 
distributions respectively. The table models 
consist of the following parameters: temperature of the inner boundary of the 
accretion disk ($T_{in}$), thermal electron temperature ($KT$) or the electron powerlaw index ($\Gamma_e$) , size of the 
corona ($Size$), scattering optical depth of the corona ($\tau$), inclination angle of the accretion disk, and 
normalization parameter ($K_{BB}$) which is added by the $XSPEC$ 
automatically. 

\begin{figure}
\centerline{ \vbox{ \centerline{ \hbox{ 
\psfig{figure=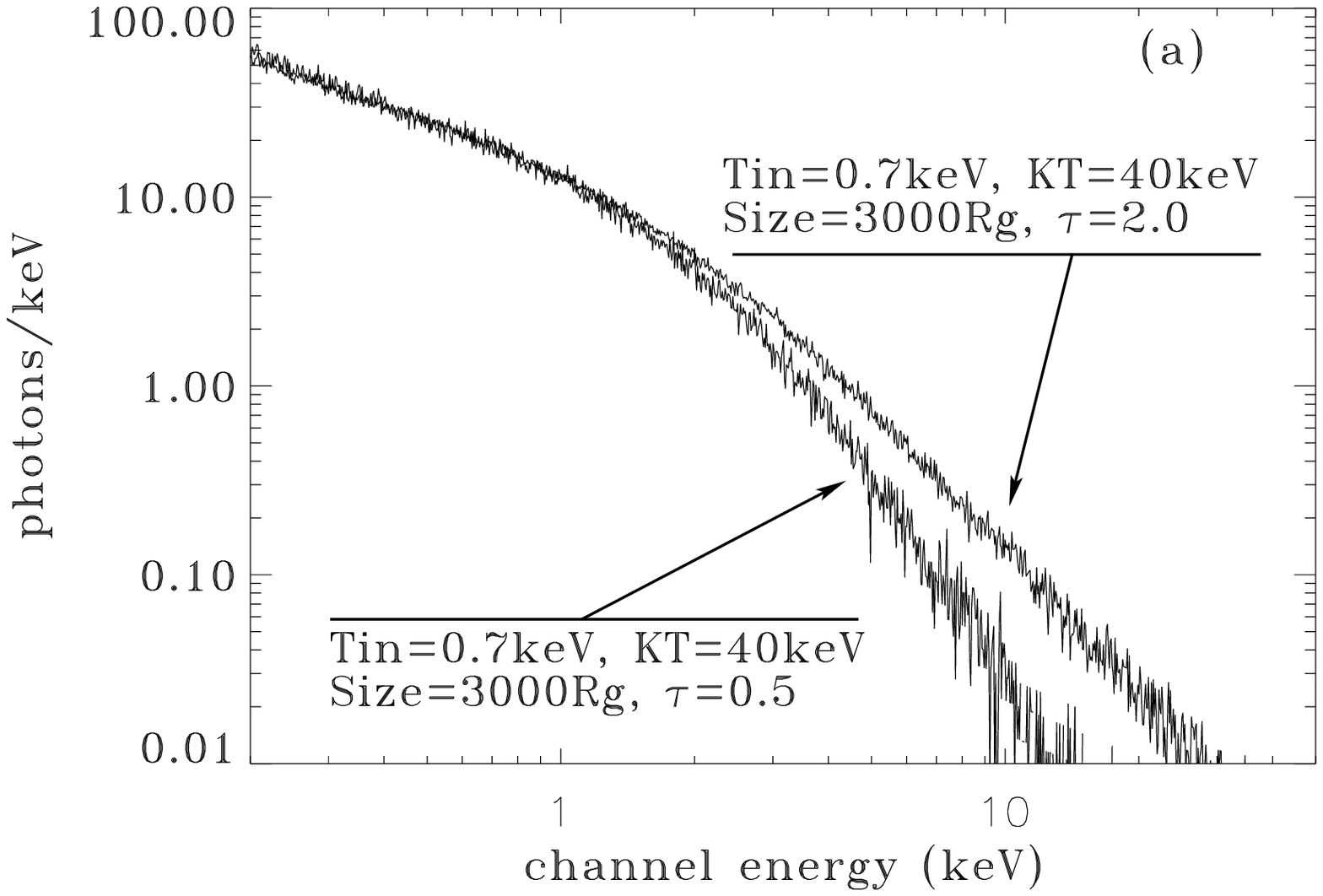,width=2.8in} 
\psfig{figure=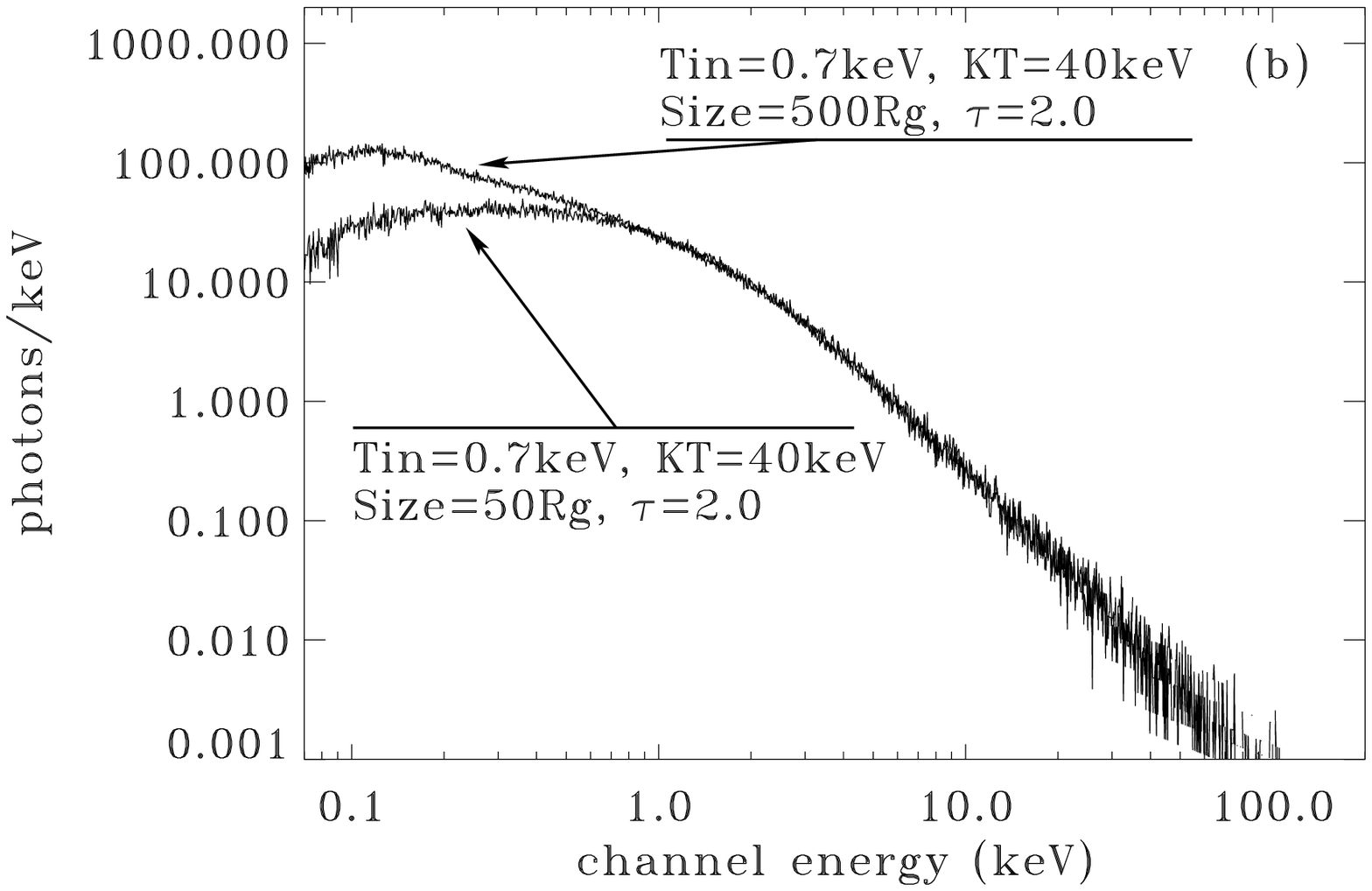,width=2.8in} }} \centerline{ \hbox{ 
\psfig{figure=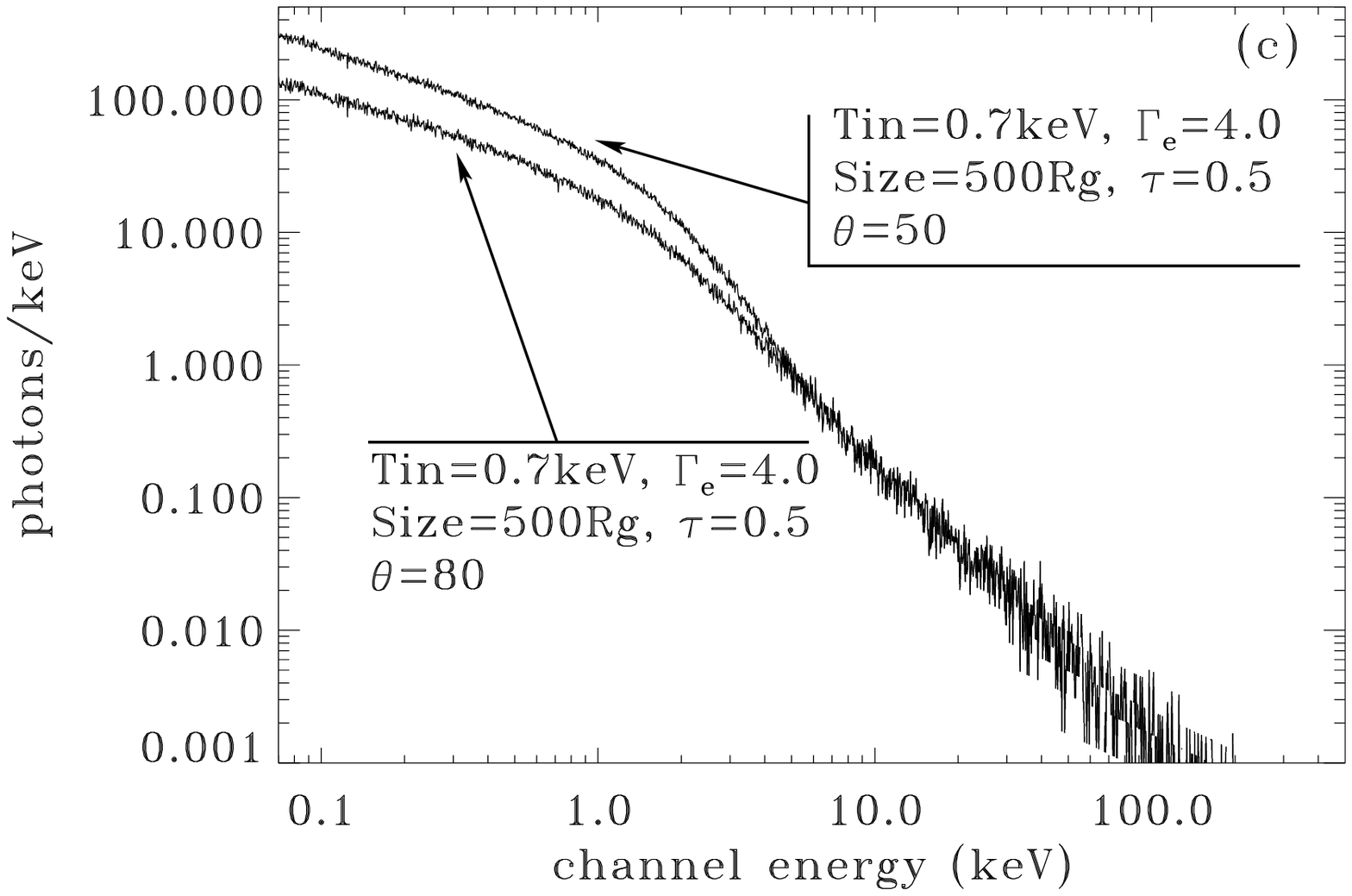,width=2.8in} 
\psfig{figure=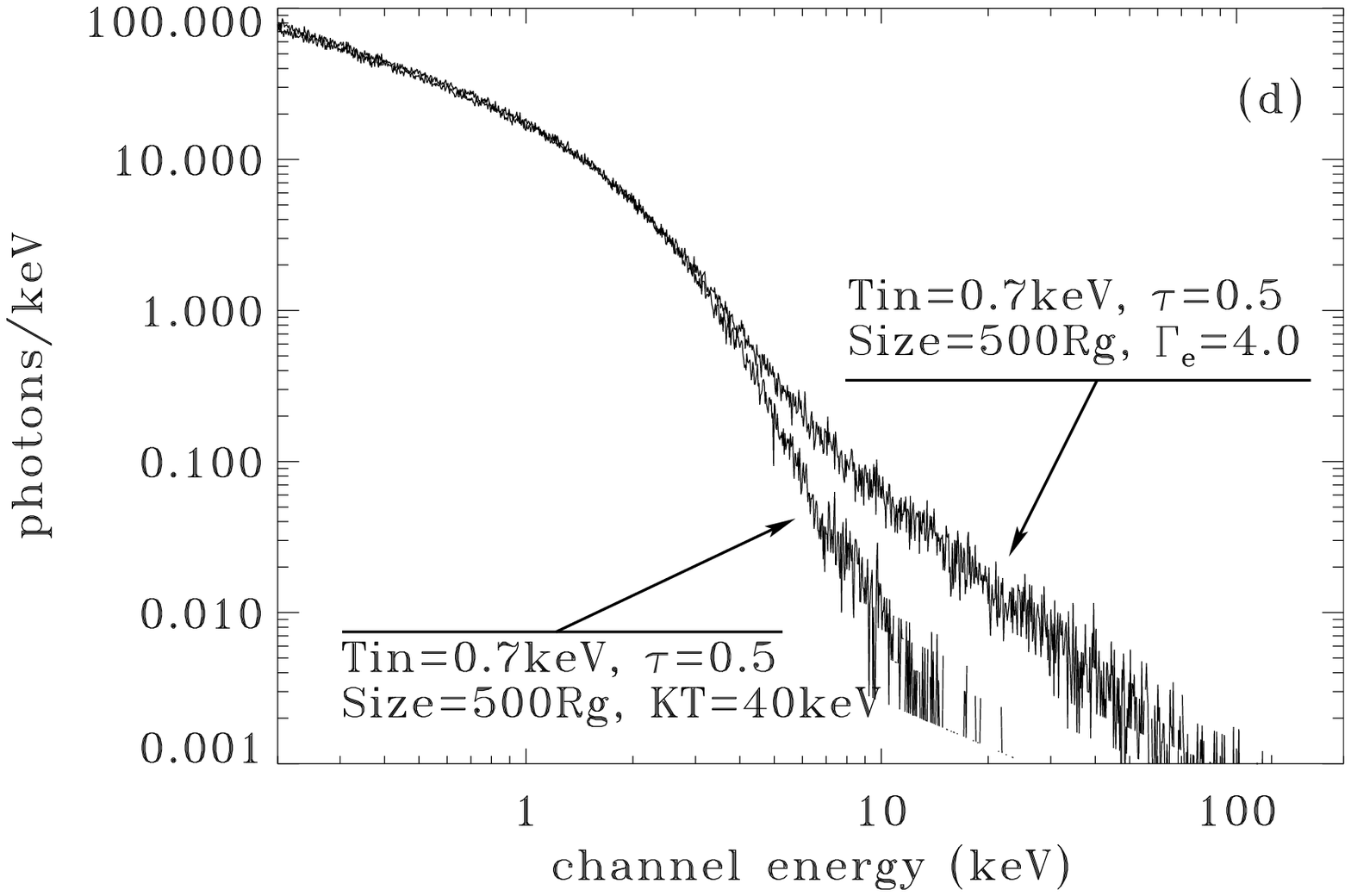,width=2.8in} }} }} \caption{\small X-ray 
photon spectra for the spherical corona with different physical properties 
of the accretion disk and the corona. Panel (a) and panel (b) show the 
photon spectra for the thermal electron energy distribution with a 
temperature of 40 keV. Panel (a) is for a corona with different optical 
depths but with the same size. Panel (b) is for a corona with different 
sizes but with the same optical depth. Panel (c) shows the different photon 
spectra with different inclination angles when the electron energy 
distribution has the powerlaw form. Panel (d) shows the comparison of the 
photon spectra between thermal and powerlaw electron energy distributions 
in the corona.} \label{therm} 
\end{figure}

\section{Application to XTE J2012+381}
\vspace{-3mm} The two models are applied to model the 
BeppoSAX broad-band data on the black hole candidate XTE J2012+381.  
FIGURE 2 shows the modeling results with $diskbb+powerlaw$ model in $XSPEC$ 
and the results with our table models. During the rising phase of the 
outburst, the powerlaw component was very strong (Sun et al., 2001)(FIGURE 
2(a)). For the BeppoSAX observation during the rising phase we have 
analyzed, the observed spectrum can be fitted reasonably well using $diskbb 
+ powerlaw$ model with photon index around 2.4 (FIGURE 2(b)). When fitting 
with our table model for the thermal electron case, the fit to the high energy data above 10 keV 
is not acceptable (FIGURE 2(c)) and thus the thermal model is rejected. However, when fitting with our table 
model for the powerlaw electron distribution, the result is comparable to the 
that with the $diskbb + powerlaw$ model. The powerlaw index for electron 
energy distribution is around 4, the size of corona is around several 
tens of gravitational radius and the inclination angle of the disk is 
around 60 degrees (FIGURE 2(d)). 
\begin{figure}
\centerline{ \vbox{ \centerline{ \hbox{ \psfig{figure=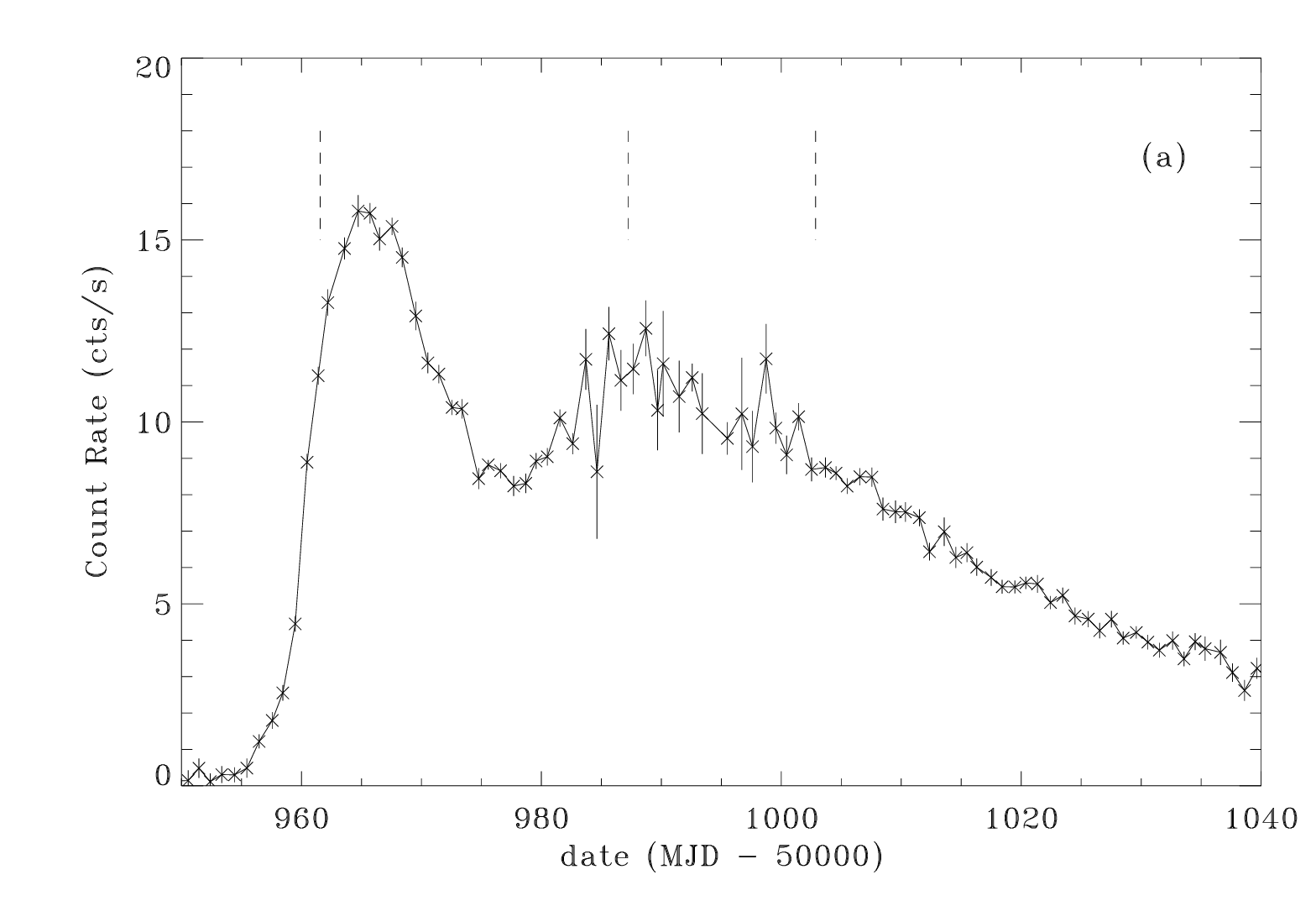,width=3.1in} 
\psfig{file=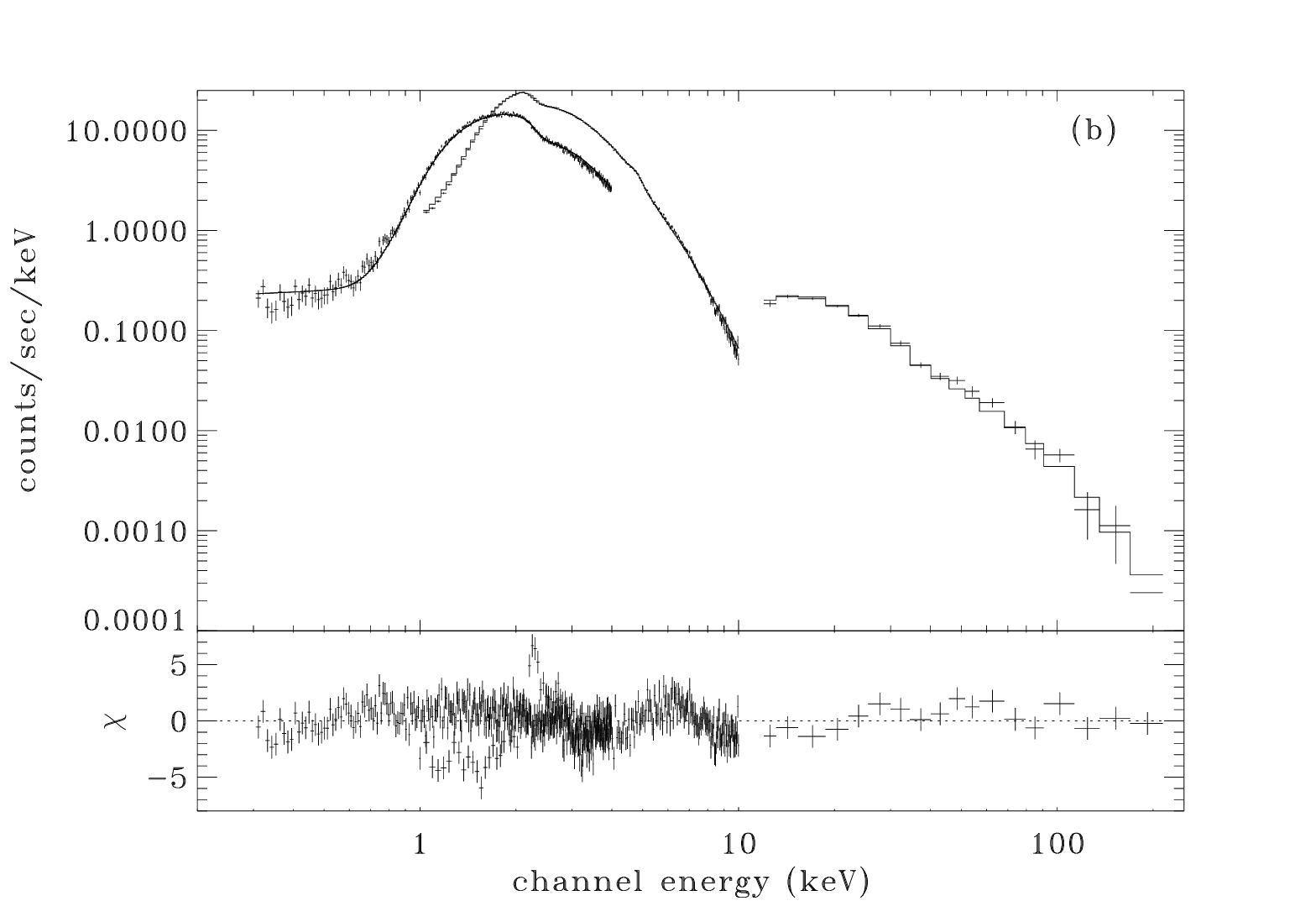,width=3.1in} }} \centerline{ \hbox{ 
\psfig{figure=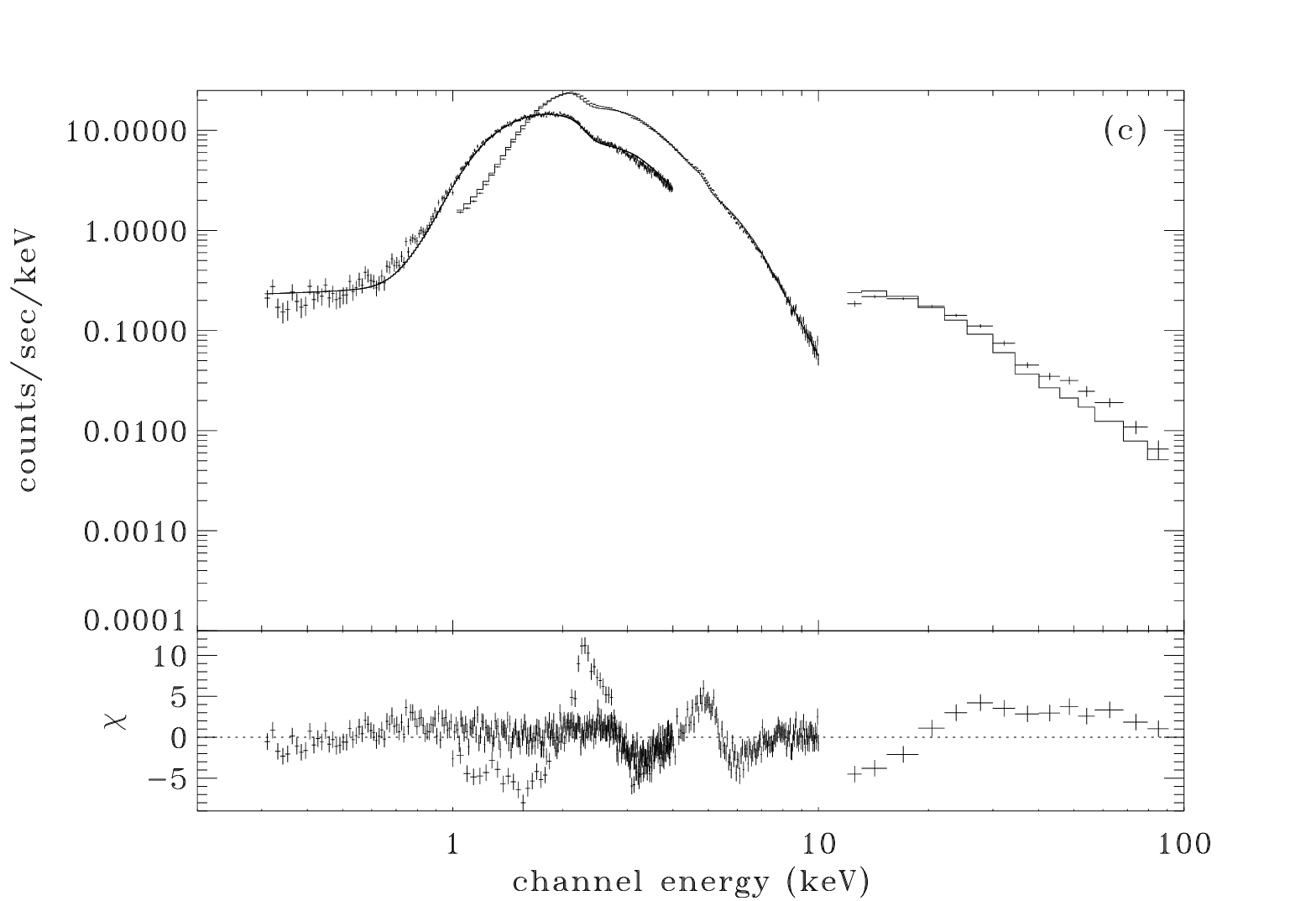,width=3.1in} 
\psfig{figure=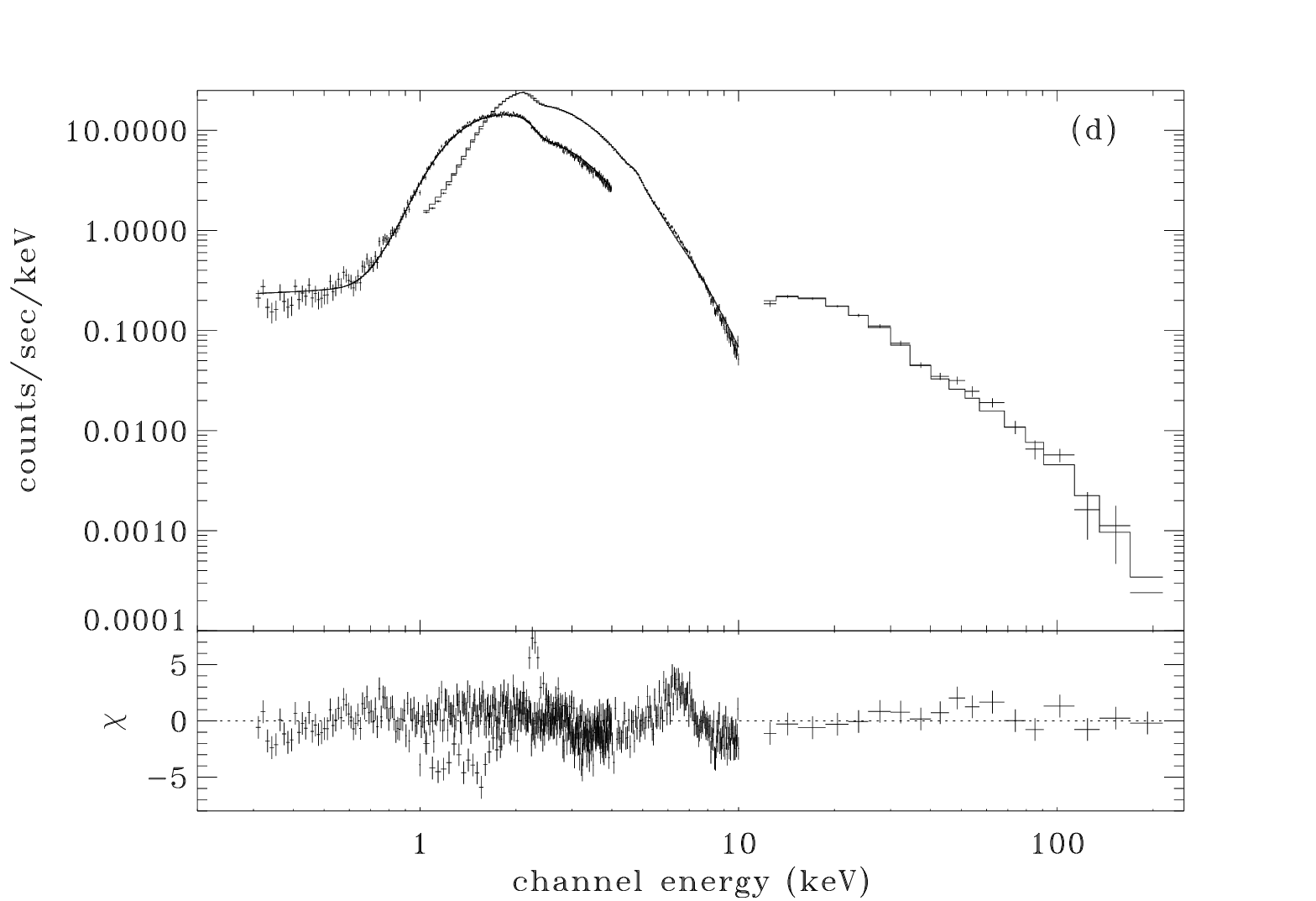,width=3.1in} }} }} \caption{\small Panel 
(a) shows the ASM light curve of the black hole candidate XTE J2012+381 
during its burst in 1998 and the dotted lines in the plot indicate the 
times of BeppoSAX observations we have analyzed. The first observation is 
in the rising phase in which the hard component is relatively strong and 
the other two observations are in the soft state in which the hard component 
is very weak. Panel (b) shows the modeling result for the first observation 
with $diskbb+powerlaw$ model. The best fit parameters are: 
$T_{in}=0.77\pm0.002\ keV,\ K_{BB} = 887\pm10.0,\ \Gamma=2.4\pm0.03$, and 
the Chi-Squared is 1368 with 553 degrees of freedom. Panel (c) shows the 
modeling result with our table model for thermal electron distribution in 
the corona. The best fit parameters are: $T_{in} = 0.6 \pm 0.02\ keV,\ KT = 
52\pm 6.2\ keV, Size = 19\pm1.7 R_g,\ \tau = 0.6\pm0.06$, and the 
Chi-Squared is 3109 with 548 degrees of freedom. Panel (d) shows the 
modeling result with our table model for powerlaw electron distribution 
 in the corona. The best fit parameters are: $T_{in} = 0.76\pm0.002\ keV,\ \Gamma_e 
=4,\ Size = 49\pm6.2R_g,\ \tau = 0.09\pm 0.003,\ \theta = 59^0\pm 3.6^0$, 
and the Chi-Squared is 1564 with 552 degrees of freedom.} \label{therm} 
\end{figure}

FIGURE 3 shows the results of inner disk radius inferred with different 
models. When modeling with $diskbb+powerlaw$ model and $diskbb+compTT$ 
model, the inner disk radius inferred directly from the normalization of 
$diskbb$ model are very small compared to the results obtained with our 
table model. The results after the radiative transfer correction (see Zhang 
et al. 2001 for detail) are consistent with the results of our table model 
reasonably well. 

\begin{figure}
\begin{minipage}{0.65\textwidth}
\psfig{figure=radius.epsi,width=2.8in} 
\end{minipage}
\begin{minipage}{0.01\textwidth}
\hspace{0.001mm} 
\end{minipage}
\begin{minipage}{0.48\textwidth}
{\bf \small FIGURE 3.} {\small Inner disk radius ($R_{in}^2 \propto 
K_{BB}$) inferred with different models. These three data points correspond 
to the three observations indicated in FIGURE 2(a). When modeling with 
$diskbb+powerlaw$ model and $diskbb+compTT$ model, the results inferred 
directly from the soft component are quite small compared to the results 
inferred from our table model and the results after the radiative transfer 
correction. } 
\end{minipage}
\label{therm} 
\end{figure}

\section{conclusion and discussion}
\vspace{-3mm} According to our simulation results and data analysis, the 
X-ray spectral shape in the low energy band might be related to the size of 
corona and the electron energy distribution may be inferred from the 
spectral shape in the high energy band. Two powerful table models have been 
built up based on our simulation results and the physical parameters can be 
obtained directly when modeling the data with these table models. 

According to the fitting results with our table models, the electron energy 
distribution in the corona of XTE J2012+381 during the rising phase seems 
to have a powerlaw form rather than the thermal form and the size of the 
corona is just several tens of gravitational radii and the inclination 
angle of the disk is around 60 degrees. 

The reason that the size of the corona may be determined from the X-ray 
data is because in our model the seed photons for the inverse 
Compton scattering come from the disk, whose temperature is a function of the distance from the central 
black hole. The relatively small size of the corona indicates that most of the hard X-ray photons come from
the region very close to the black hole. 

Another interesting result is that the spectral fitting is quite sensitive 
to the disk inclination angle. This is because the observed flux from the 
disk depends strongly upon the inclination angle, while the hard X-ray 
flux is almost isotropic, especially for the case of spherical corona. 
Unfortunately the inclination angle and the geometrical factor (opaqueness 
of the corona) may be coupled together in the spectral fitting if the 
quality of the data is not good enough. If the 
inclination angle of the system is determined independently, e.g., with 
optical photometry observations or radio measurements of jets, the shape of the corona may be constrained reliably 
with X-ray spectral fitting.

Determining the 
value of the inner disk radius is very important in 
understanding the physics of the accretion disk and the black hole angular momentum (Zhang, 
Cui and Chen 1997). The normalization parameter inferred with our table model, which is proportional to the square of the 
inner disk radius (Makishima et al., 1986), is significantly different 
from that determined with the simple $diskbb+powerlaw$ or $diskbb+compTT$ 
model without radiative transfer correction. This is because 
the corona, though optically thin in most cases, scatters some of the photons emitted from the disk and make the 
observed soft component different
significantly from the original disk emission. 

Therefore modeling broad-band X-ray continuum spectra with physically consistent and accurate models provides
a powerful tool in determining the properties of accretion disks and coronae in black hole X-ray 
binaries. However such studies require high quality and broad-band data, which 
may be provided by BeppoSAX, Chandra and XMM currently. Future data from 
Integral, Swift and especially the Constellation X-ray missions will provide 
significant breakthroughs in this field.

Currently, we have not included non-uniform corona and other geometry in 
our table models. The general 
relativistic effects and Doppler effects (Zhang, et al., these proceedings) are also not taken into account in our 
table model currently. All of these will be our future work. 

{\bf Acknowledgments:} Mr. Yongzhong Chen is acknowledged 
for his initial work on this project during his visit to UAH in 1999-2000. We thank
Drs. Lev Titarchuk, Wei Cui and Yuxin Feng for interesting discussions. This work was supported in part by NASA Marshall Space 
Flight Center under contract NCC8-200 and by NASA Long Term Space 
Astrophysics Program under grants NAG5-7927 and NAG5-8523. 


\end{document}